\documentclass[sigconf]{acmart}

\usepackage{multirow}
\usepackage{graphicx}
\usepackage[normalem]{ulem}
\usepackage{diagbox}
\usepackage{makecell}
\usepackage{bm}
\usepackage{xspace}
\usepackage{url}
\usepackage{array}
\usepackage{verbatim}
\usepackage{soul, color}
\usepackage{enumitem}

\AtBeginDocument{%
	\providecommand\BibTeX{{%
			\normalfont B\kern-0.5em{\scshape i\kern-0.25em b}\kern-0.8em\TeX}}}

\definecolor{ao}{rgb}{1.0, 0.73, 0.0}

\newcommand{\eat}[1]{}

\newcommand{\ie}{\emph{i.e.,}\xspace}
\newcommand{\eg}{\emph{e.g.,}\xspace}

\newcommand{\paratitle}[1]{\vspace{0.5em}\noindent\textbf{#1}}

\newcommand{\hlgrn}[1]{{\sethlcolor{ao}\hl{#1}}}
\setul{}{1pt}
\newcommand{\ulred}[1]{{\setulcolor{red}\ul{#1}}}

\copyrightyear{2020} 
\acmYear{2020} 
\setcopyright{acmcopyright}\acmConference[SIGIR '20]{Proceedings of the 43rd International ACM SIGIR Conference on Research and Development in Information Retrieval}{July 25--30, 2020}{Virtual Event, China}
\acmBooktitle{Proceedings of the 43rd International ACM SIGIR Conference on Research and Development in Information Retrieval (SIGIR '20), July 25--30, 2020, Virtual Event, China}
\acmPrice{15.00}
\acmDOI{10.1145/3397271.3401169}
\acmISBN{978-1-4503-8016-4/20/07}

\title{CATN: Cross-Domain Recommendation for Cold-Start Users via Aspect Transfer Network}\thanks{$^\star$Chenliang Li is the corresponding author.}
\author{Cheng Zhao$^1$, Chenliang Li$^{2\star}$, Rong Xiao$^3$, Hongbo Deng$^3$, Aixin Sun$^4$}
\affiliation{
	\institution{
		1. State Key Laboratory of Information Engineering in Surveying, Mapping and Remote Sensing, Wuhan University, Wuhan, China\\
		2. School of Cyber Science and Engineering, Wuhan University, Wuhan, China\\
		3. Alibaba Group, Hangzhou, China\\
		4. School of Computer Science and Engineering, Nanyang Technological University, Singapore
}}
\email{{zhaocheng\_whuer,cllee}@whu.edu.cn, xiaorong.xr@taobao.com, dhb167148@alibaba-inc.com, axsun@ntu.edu.sg}
\def\authors{Cheng Zhao, Chenliang Li, Rong Xiao, Hongbo Deng, Aixin Sun}

\begin{document}
	\fancyhead{}
	
	\settopmatter{printfolios=false}
	
	\begin{abstract}		
In a large recommender system, the products (or items) could be in many different categories or domains. Given two relevant domains (\eg \textit{Book} and \textit{Movie}), users may have interactions with items in one domain but not in the other domain. To the latter, these users are considered as cold-start users. How to effectively transfer users' preferences based on their interactions from one domain to the other relevant domain, is the key issue in cross-domain recommendation. Inspired by the advances made in review-based recommendation, we propose to model user preference transfer at aspect-level derived from reviews. To this end, we propose a \textbf{c}ross-domain recommendation framework via \textbf{a}spect \textbf{t}ransfer \textbf{n}etwork for cold-start users (named CATN).  CATN is devised to extract multiple aspects for each user and each item from their review documents, and learn aspect correlations across domains with an attention mechanism. In addition, we further exploit auxiliary reviews from  like-minded users to enhance a user's aspect representations. Then, an end-to-end optimization framework is utilized to strengthen the robustness of our model. On real-world datasets, the proposed CATN outperforms SOTA models significantly in terms of rating prediction accuracy. Further analysis shows that our model is able to reveal user aspect connections across domains at a fine level of granularity, making the recommendation explainable.
\end{abstract}

	\maketitle
		
\section{Introduction}\label{sec:intro}
Recommender systems play vital roles in various e-commerce platforms. Traditional collaborative filtering methods recommend items to users mainly based on their historical feedbacks. However, these approaches become less effective for new users, \ie cold-start users, who have no historical feedbacks. Recently, cross-domain recommendation has gained wide attention~\cite{ijcai11:cdr,ijcai18:dcdsr}. Given two relevant domains (\eg \textit{Book} and \textit{Movie}), users may have historical interactions in one domain (\ie source domain), but not the other (\ie target domain). To the target domain, these users are considered as cold-start users. However, as the two domains are relevant, feedbacks in the source domain could be leveraged to provide meaningful recommendations in target domain.

\begin{figure}
		\centering
		\includegraphics[width=0.96\linewidth]{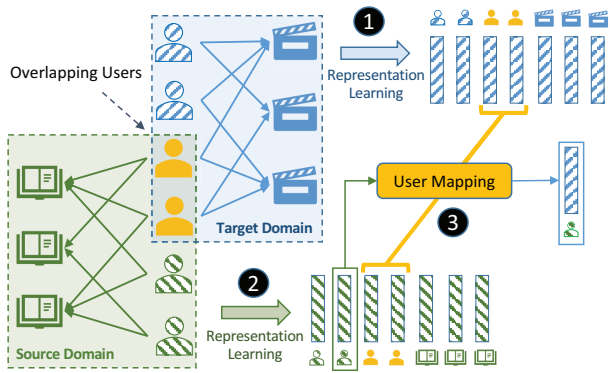}
		\caption{Existing workflow in cross-domain recommendation for cold-start users (\textit{Best viewed in color}).}\label{fig:3-step}
	\end{figure}

The core task of cross-domain recommendation is user preference mapping between the two relevant domains. To achieve the mapping,  existing approaches such as EMCDR~\cite{ijcai17:emcdr}, CDLFM~\cite{dasfaa18:cdlfm} and RC-DFM~\cite{aaai19:rc_dfm} encode users' preference into single vectors, then conduct cross-domain mapping as a whole. Illustrated in Figure~\ref{fig:3-step}, existing solutions learn user/item representations in source domain and target domain respectively. Then, cross-domain representation mapping is learned based on the overlapping users. Note that, the direct mapping between user representations of source and target domains cannot explicitly capture users' diverse yet fine-grained preferences in different domains. For example, a user who prefers Chinese kung fu novels is more likely to be fund of Chinese ancient dramas. \eat{a user who prefers detective novels is more likely to be fond of movies with complex plots, in contrast to movies with only special effects.}

In our study, we assume users' preferences are multi-faceted, \eg \textit{plot}, \textit{text description}, \textit{scene} in \textit{Book} and \textit{Movie} domains. Modeling these fine-grained semantic aspects and exploring their mutual relationships across domains, would lead to better user preference understanding and explainable recommendation. To this end, we aim to exploit user/item reviews for cross-domain aspect correlations. In recent years, there has been a surge of approaches utilizing user/item reviews for aspect-based recommendation (\ie rating prediction for a given user-item pair)~\cite{kdd14:jmars,www18:aflm,cikm18:anr,sigir19:carp}. Inspired by their encouraging performance, we propose to explore users' preferences based on the aspects generated from reviews across domains.

\eat{
Lastly, we replace the defective three-step optimization process with a novel end-to-end strategy, which consists of two training flows derived from two domains, thus strengthening the robustness of training process.
}
	
In this paper, we propose a \textbf{c}ross-domain recommendation framework for cold-start users via \textbf{a}spect \textbf{t}ransfer \textbf{n}etwork, named CATN. In source domain, we represent a user by a \textit{user document} which contains all reviews written by this user, and an item by an \textit{item document} which contains all reviews it receives. The same applies in target domain. An overlapping user therefore will have two user documents, one in source domain and the other in target domain. To extract aspects mentioned in user and item documents, we utilize an aspect-specific gate mechanism over a convolutional layer. Then, global cross-domain aspect correlations are identified and weighted through attention mechanism, for preference estimation. To support review-based knowledge transfer, we introduce a novel cross-domain review-based preference matching procedure with two learning flows. The illustration of these two learning flows is shown in Figure~\ref{fig:flows}. Specifically, for a given overlapping user and an item in the target domain, we utilize the user's review document in source domain and the item's review document in target domain  to perform rating prediction, and vice versa. These two learning flows are proceeded in turn with the guidance of the global cross-domain aspect correlations. Considering review scarcity~\cite{cikm18:parl} and the small number of overlapping users~\cite{cikm19:sscdr}, we further enhance user representation by an additional \textit{user auxiliary document} for each user. An auxiliary document contains all reviews written by the like-minded users, \ie the users who give the same rating to the same item as the current user. The auxiliary documents are also utilized in aspect extraction.

We summarize our key contributions as follows. We propose a novel deep recommendation model for cold-start users, by bridging multiple user's inherent traits via reviews in different domains. To the best of our knowledge, this is the first attempt to learn cross-domain aspect-level preference matching, in an end-to-end learning fashion. Through extensive experiments conducted on three pairs of real-world datasets, we demonstrate that CATN performs significantly better than state-of-the-art (SOTA) alternatives. We also conduct detailed analysis to validate the benefit introduced by each component of CATN, and show how CATN works at a fine-grained semantic level.
	
\section{Related Work}
Our work is related to two subareas of recommender systems: cross-domain recommendation, and aspect-based recommendation. Next, we briefly review the works in each subarea.
\eat{
Next, we briefly review the works in two subares: cross-domain recommendation and aspect-based recommendation.
}

\subsection{Cross-Domain Recommendation}
\eat{As an important branch of recommender system, cross-domain recommendation has been widely studied recently.} By leveraging relevant source domain as auxiliary information, a surge of solutions are proposed to address the data sparsity and cold-start problems for the target domain. At the very beginning, CMF~\cite{kdd08:cmf} proposes to achieve knowledge integration across domains by concatenating multiple rating matrices and sharing user factors across domains. Then Temporal-Domain CF~\cite{ijcai11:cdr} shares the static group-level rating matrix across temporal domains. Later, CDTF~\cite{www13:cdtf} is proposed to capture the triadic relation of user-item-domain by tensor factorization. These collaborative filtering based works suffers severely from the data sparsity problem when considering different domains as a whole.

In recent years, with the revival of deep learning techniques, many deep learning-based models are proposed to enhance knowledge transfer.  EMCDR~\cite{ijcai17:emcdr} explicitly maps user representations from different domains via a multi-layer fully connected neural network. DCDCSR~\cite{ijcai18:dcdsr} further extends EMCDR by generating benchmark factors to solve cross-domain and cross-system problems. CoNet~\cite{cikm18:conet} is proposed to train a deep cross-stitch network for enhancing the recommendation on both domains simultaneously. PPGN~\cite{cikm19:ppgn} leverages the user-item interaction graph to capture the process of user preference propagation. DARec~\cite{ijcai19:darec}, equipped with an adversarial learning process, is proposed for user-item rating prediction. $\pi$-Net~\cite{sigir19:pi-net} is devised for shared-account cross-domain sequential recommendation.

To avoid the leak of user privacy, NATR~\cite{www19:natr} chooses to transfer only the item embeddings across domains. SSCDR~\cite{cikm19:sscdr} investigates the distribution of cross-domain overlapping users in real-world scenarios, and come up with a semi-supervised mapping approach to perform recommendation for cold-start users. CDLFM~\cite{dasfaa18:cdlfm} modifies the matrix factorization and mapping process by exploiting the user neighborhoods. Another line of cross-domain recommender systems is clustering-based, which has also achieved good performance.  $C^3R$~\cite{sigir17:c3r} leverages users' multiple social media sources to boost the performance of venue recommendation. CDIE-C~\cite{wsdm19:clustercr} enhances item embedding learning by means of cross-domain co-clustering. 
	
Nevertheless, many of the above solutions only consider rating records while ignoring other complementary yet fertile information, \eg reviews. MVDNN~\cite{www15:mvdnn} maps users' and items' auxiliary information to a latent space where the similarity between users and their preferred items is maximized. To combine the strength from both ratings and reviews, RB-JTF~\cite{dasfaa17:cdrjtf} transfers users' preference by a joint tensor factorization derived from the reviews. RC-DFM~\cite{aaai19:rc_dfm} trains user or item factors with a review-fused SDAE, which achieves the SOTA performance for cold-start user recommendation. 

Existing review-based transfer solutions have earned substantial improvement over traditional interaction-based methods. However, these works still have many drawbacks to be overcome. As discussed in Section~\ref{sec:intro} and illustrated in Figure~\ref{fig:3-step}. Existing solutions learn users and items representations in  source domain and target domain respectively (steps 1 and 2 in Figure~\ref{fig:3-step}). Then they learn the cross-domain representation mapping based on the overlapping users (step 3 in Figure~\ref{fig:3-step}). This mapping cannot explicitly distinguish the fine-grained semantic characteristics. Further,  the pipelined learning process could easily accumulate and magnify noisy information produced by the sub-optimal learning in the intermediate steps. We therefore propose a completely different network architecture, to capture and align the fine-grained user preferences between source and target domains at aspect level, through reviews, and in an end-to-end fashion. 
	
\subsection{Aspect-based Recommendation}
Reviews reflect a user's purchased experience, and have shown to be effective in addressing the sparsity problem in recommendation. Nowadays, review-based recommender systems have become a pivotal building block for recommendation in single-domain~\cite{wsdm17:deepconn,recsys17:d-attn,cikm18:anr,kdd18:mpcn,kdd19:liu,tois19:carl,sigir19:carp,cikm19:xia,sigir20:rrs}. Within review-based recommender systems, aspect-based recommender systems, which model the fine-grained relations between user preferences and item characteristics, have drawn great attention recently. In general, existing solutions for aspect-based recommender systems can be divided into two main categories.

Solutions in the first category extract aspects and sentiments from reviews by utilizing external NLP toolkits. Example solutions include MTER~\cite{sigir14:mter}, TriRank~\cite{cikm15:trirank}, LRPPM~\cite{sigir16:lrppm}, SULM~\cite{kdd17:sulm} and EFM~\cite{sigir18:efm}. The performance of such solutions therefore are highly dependent on the quality of the external toolkits used in the process.
	
The second category of solutions fulfills automatic aspects extraction, with an internal model component. For example, JMARS~\cite{kdd14:jmars} utilizes topic modeling to learn multiple aspect representations. Following JMARS, FLAME~\cite{wsdm15:flame} and AFLM~\cite{www18:aflm} are proposed to model aspect-level user preferences and item characteristics through an integrated hidden topic learning process. However, the static representations learned by these methods are incapable of modeling dynamic and complex relationships between users and items. To dynamically model the relation encoded by different user-item pairs, ANR~\cite{cikm18:anr} uses a co-attention mechanism to infer the importance of different aspects with respect to a given user-item pair. More recently, CARP~\cite{sigir19:carp} proposes a capsule network to conduct rating prediction and provide interpretability in a fine-grained manner.
	
Note that, solutions in both categories focus on the single-domain recommendation. These methods cannot handle the cold-start user whose historical interactions are not available in the target domain. In this work, we make the first attempt to complete this picture by designing a cross-domain aspect transfer network to achieve recommendation for cold-start users in target domain.

%
%

\section{The CATN Framework}
\eat{In this section, we present our proposed CATN framework in detail.} We start with the problem setting of cross-domain recommendation for cold-start users. Then, we provide an overview of CATN along with the motivation behind its each component. After presenting all the components, we go through the optimization process.

\subsection{Problem Formulation}

\eat{For convenience, the key notations used throughout this paper are summarized in Tab.~\ref{tab:notions}.}

We use $\mathcal{D}_s$ and  $\mathcal{D}_t$ to denote source domain and target domain respectively. Note that a domain includes its users, items, and interactions (\eg ratings and reviews) between users and items. Let $U^{o}$ be the set of overlapping users,  who have historical interactions with items in both $\mathcal{D}_s$ and $\mathcal{D}_t$. $U^{cs}$ denotes the set of cold-start users who have interactions with items in  $\mathcal{D}_s$, but not with items in $\mathcal{D}_t$. For a given cold-start user $u\in U^{cs}$, our task is to estimate the rating $\hat{r}_{u,i}$ that user $u$ would give to an item $i$ in $\mathcal{D}_t$.

	\eat{
	\begin{table}
		\centering
		\caption{List of Notations.}
		\setlength{\tabcolsep}{1mm}{
			\scalebox{0.9}{
				\begin{tabular}{cc}
					\toprule[1.2pt]
					Notation & Definition \\
					\midrule[1pt]
					$\mathcal{D}_s$ & source domain \\
					$\mathcal{D}_t$ & target domain \\
					$r_{u,i}$ & the rating from user $u$ to item $i$ \\
					$D_{u}$ & \makecell{the user document of user $u$} \\
					$D_{u_{aux}}$ & \makecell{the user auxiliary document of user $u$} \\
					$D_{i}$ & \makecell{the item document of item $i$} \\
					\midrule
					$M$ & \makecell{number of aspects extracted from\\ user/item document} \\
					$k$ & latent dimension of aspect representation\\
					$\mathbf{v}_s^m, \mathbf{v}_t^m \in \mathbb{R}^k $ & \makecell{global-sharing aspect representations of\\ $\mathcal{D}_s$, $\mathcal{D}_t$, $m=1,..,M$} \\
					\bottomrule[1.2pt] \label{tab:notions}
		\end{tabular}}}
		
	\end{table}
	}
	
\subsection{Overview of CATN}	
\begin{figure*}[t]
	\centering
	\includegraphics[width=0.90\linewidth]{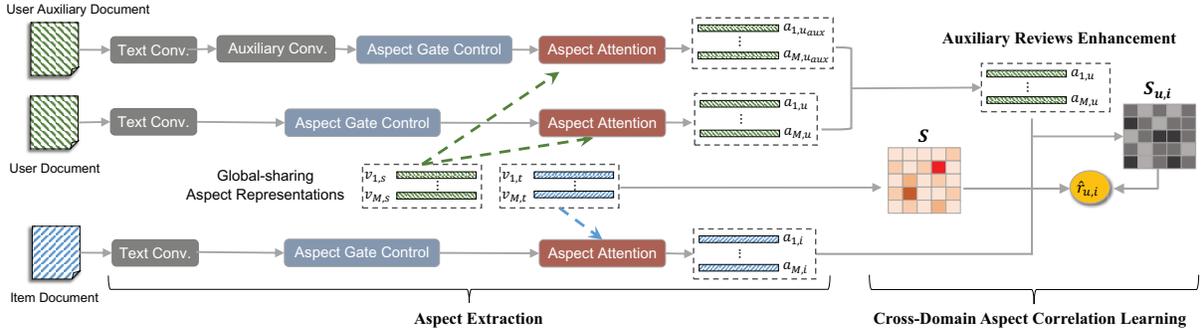}
	\caption{The architecture of CATN (\textit{Best viewed in color}).}\label{fig:architecture}
\end{figure*}

The overall structure of CATN is illustrated in Figure~\ref{fig:architecture}. Its structure consists of three components:  \textit{Aspect Extraction}, \textit{Auxiliary Reviews Enhancement}, and \textit{Cross-Domain Aspect Correlation Learning}. Because our task is to achieve review-based cross-domain preference transfer, the overall procedure for rating prediction differs fundamentally from the existing review-based recommendation systems in single-domain. Here, the ratings and the reviews of the overlapping users $U^{o}$ from both $\mathcal{D}_s$ and $\mathcal{D}_t$ are used for model training.

\begin{figure}[t]
	\centering
	\includegraphics[height=40mm,width=0.68\linewidth]{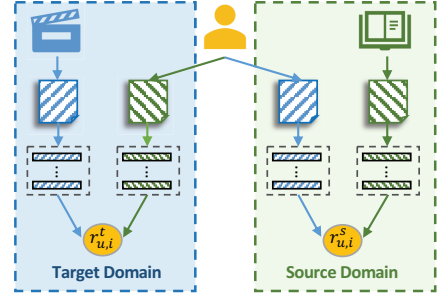}
	\caption{Two learning flows in CATN (\textit{Best viewed in color}).}\label{fig:flows}
\end{figure}

In source domain, we represent a user by a \textit{user document} $D_u$, and an item by an \textit{item document} $D_i$. Similarly, each user and each item in the target domain has a user  document and an item  document, respectively. An overlapping user will have two user documents, one from source domain $D_u^s$ and the other from target domain $D_u^t$. We use superscript ``$s$'' and ``$t$'' to indicate the source domain and target domain for a clear presentation. Recall that an overlapping user has interactions with items in both source and target domains. For a given overlapping user $u\in U^o$, as shown in Figure~\ref{fig:flows}, we devise a cross-domain review-based preference matching procedure with two learning flows: 1) her user  document $D_u^s$ in source domain and the item document $D_i^t$ for item $i$ in target domain are utilized in model training to match $r_{u,i}^t$ in target domain; and 2) her user  document $D_u^t$ in target domain and the item  document $D_i^s$ of an item $i$ in source domain are utilized to match $r_{u,i}^s$ in source domain.

The matching of user preference between source and target domains are achieved at \textit{aspect level}, derived from the user and item  documents, as shown in Figure~\ref{fig:architecture}. Note that, in addition to user document, we also utilize an auxiliary review document for each user. This auxiliary review document contains reviews written by like-minded users, to be detailed shortly. Aspects derived from the two kinds of user documents are merged. Then, a cross-domain aspect correlation learning will distinguish the more correlated aspect-pair across domains to conduct rating predictions. Next, we detail the aspect extraction process.


\subsection{Aspect Extraction}\label{ssec:aspectExt}
To extract aspects, the same process is applied to user document $D_u$ and item document $D_i$, in both source and target domains. As the procedure is the same, we take $D_u$ as a running example.
	
\paratitle{Text Convolution}. Given a user  document $D_{u}=[w_1,w_2,..,w_l]$, we first project each word to its embedding representation: $\mathbf{E}_{u}=[\mathbf{e}_1,\mathbf{e}_2,..,\mathbf{e}_l]$, $ \mathbf{e}_j\in \mathbb{R}^d$, where $l$ is the document length and $d$ is the word embedding dimension. In order to capture the context information around each word, we perform a convolution operation with an $ReLU$ activation function. Here, $n$ convolution filters with the same sliding window of size $s$ are applied over matrix $\mathbf{E}_{u}$ to extract contextual features. The resultant feature matrix is $\mathbf{C}_{u}=[\mathbf{c}_{1,u},\mathbf{c}_{2,u},..,\mathbf{c}_{l,u}]$, where $\mathbf{c}_{j,u}\in \mathbb{R}^n$ is the latent contextual feature vector for $j$-th word.

\eat{
Specifically, for word $w_h$, each filter $f_i$ performs a convolutional operation as follows:
		\begin{equation}
	c_{h,u}^i = ReLU(\mathbf{W}_{c}^i * \mathbf{E}_{u}[h-\frac{s-1}{2}:h+\frac{s-1}{2}] + b_{c}^i)\label{eqn:textConv}
	\end{equation}
In this equation, $*$ is the convolution operator. $\mathbf{W}_{c}^i \in \mathbb{R}^{s \times d} $ is the weight matrix for filter $f_i$. $\mathbf{E}_{u}[h-\frac{s-1}{2}:h+\frac{s-1}{2}]$ is the slice of matrix $\mathbf{E}_{u}$ within the sliding window centering at position $h$. $ {b}_{c}^i$ is the bias term, and $c_{h,u}^i$ is the local contextual features extracted by filter $f_i$ over the sliding window for word $w_h$. By concatenating $c_{h,u}^i$ in rows and columns, we get the resultant feature matrix $\mathbf{C}_{u}=[\mathbf{c}_{1,u},\mathbf{c}_{2,u},..,\mathbf{c}_{l,u}]$, $\mathbf{c}_{j,u}\in \mathbb{R}^n$ is the latent contextual feature vector for $j$-th word.
}
	
\paratitle{Aspect Gate Control}. The contextual features $\mathbf{c}_{j,u}$ extracted for $j$-th word can be considered as a composition of multiple semantic aspects. Here, we further utilize an aspect-specific gate mechanism to identify which features are relevant to each aspect. Specifically, for $m$-th aspect, the aspect-specific features $\mathbf{g}_{m,j,u}$ of word $w_j$ are extracted as follows:
	\begin{equation}
	\mathbf{g}_{m,j,u} = (\mathbf{W}_{m}\mathbf{c}_{j,u}+\mathbf{b}_{m}) \odot \sigma (\mathbf{W}_{m}^g\mathbf{c}_{j,u}+ \mathbf{b}_{m}^g)\label{eqn:aspectGate}
	\end{equation}
	where $\sigma$ is the sigmoid activation function, $\odot$ is the element-wise product operation. $\mathbf{W}_{m},\mathbf{W}_{m}^g \in \mathbb{R}^{k \times n}$ and $\mathbf{b}_{m},\mathbf{b}_{m}^g \in \mathbb{R}^{k}$ denote the transform matrices and bias vectors respectively for the $m$-th aspect. $k$ is the latent dim of aspect representation. Here, the second term on the right hand side of Equation~\ref{eqn:aspectGate} works as a soft \textit{on-off} switch controlling which latent feature is relevant to the aspect. Consequently, we get $M$ aspect-specific words contextual features $\mathbf{G_{u}}$, which are leveraged for further aspect extraction.
	\begin{align}
	\mathbf{G_{u}} &= [\mathbf{G}_{1,u}, \mathbf{G}_{2,u},..,\mathbf{G}_{M,u}], \\
	\mathbf{G}_{m,u} &= [\mathbf{g}_{m,1,u},\mathbf{g}_{m,2,u},..,\mathbf{g}_{m,l,u}]
	\end{align}
	
	\eat{
	Similarly, for item document $D_i$ of item $i$ in $\mathcal{D}_t$,  $\mathbf{G_{i_t}}$ can be generated from $D_{i_t}$ similarly.
	\begin{align}
		\mathbf{G_{i_t}} &= [\mathbf{G}_{1,i_t}, \mathbf{G}_{2,i_t},..,\mathbf{G}_{M,i_t}] \\
		\mathbf{G}_{m,i_t} &= [\mathbf{g}_{m,1,i_t},\mathbf{g}_{m,2,i_t},..,\mathbf{g}_{m,l,i_t}]
	\end{align}
	}

	\eat{Several gate linear units (GLU) are introduced to distinguish diverse aspect features carried by $\mathbf{c}_{j,u}$. $M$ is the predefined number of aspects hold by a user/item. For the $m$-th aspect, the resultant vector calculated from $\mathbf{c}_{j,u}$ can be formulated as:}
	
\paratitle{Aspect Attention}. Reviews from different domains put emphasis on different aspects. For instance, \textit{Book} domain tends to include \textit{plots} and \textit{figures}, while \textit{Movie} domain tends to include \textit{actors} and \textit{special effects}. Accordingly, we design two matrices of global-sharing aspect representations in $\mathcal{D}_s$ and $\mathcal{D}_t$. They are denoted as $\mathbf{V}_s = [\mathbf{v}_{1,s},...,\mathbf{v}_{M,s}]$ and $\mathbf{V}_t = [\mathbf{v}_{1,t},...,\mathbf{v}_{M,t}]$, for source and target domains respectively. $\mathbf{V}_s$ and $\mathbf{V}_t$ serve as the query to guide the aspect extraction. Concretely, the representation $\mathbf{a}_{m,u}$ of the $m$-th aspect extracted from $\mathbf{G}_{m,u}$ is derived as follows:
	\begin{align}
		\mathbf{a}_{m,u} &= \sum_{j=1}^{l} \beta_{m,j,u} \mathbf{g}_{m,j,u}\label{eqn:aspect}\\
		\beta_{m,j,u} &= \frac{exp(\mathbf{g}_{m,j,u}^{\top}\mathbf{v}_{m,s})}{\sum_{i=1}^{l} exp(\mathbf{g}_{m,i,u}^{\top}\mathbf{v}_{m,s})} \label{eqn:attn_s}
	\end{align}
	Here, $\beta_{m,j,u}$ indicates the importance of word $w_j$ towards the $m$-th aspect. Consequently, we can obtain the representations of $M$ aspects from $D_u$, constituting the aspect matrices $\mathbf{A}_{u}=[\mathbf{a}_{1,u},..,\mathbf{a}_{M,u}]$. Following the same precedure, we extract $M$ aspects from $D_i$: $\mathbf{A}_{i}=[\mathbf{a}_{1,i},..,\mathbf{a}_{M,i}]$. It is worthwhile to highlight that the parameters for aspect extraction for $D_u$ and $D_i$ are shared in each learning flow, though $D_u$ and $D_i$ are built with the reviews in different domains. Also a distinct set of parameters is used in each learning flow. Since we aim to map the aspect across domains, $\mathbf{V}_s$ and $\mathbf{V}_t$ are shared in their corresponding domains respectively.

	\eat{
	Meantime, the $m$-th aspect $\mathbf{a}_{m,i_t} \in \mathbb{R}^k$ calculated from $\mathbf{G}_{m,i_t}$ is generated in a similar way:
	\begin{align}
			\mathbf{a}_{m,i_t} &= \sum_{j=1}^{l} attn_{m,j,i_t} \mathbf{g}_{m,j,i_t}\\
			attn_{m,j,i_t} &= \frac{exp(\mathbf{g}_{m,j,i_t}^{\top}\mathbf{v}_{m,t})}{\sum_{i=1}^{l} exp(\mathbf{g}_{m,i,i_t}^{\top}\mathbf{v}_{m,t})} \label{eqn:attn_t}
	\end{align}
	}

\subsection{Auxiliary Reviews Enhancement}
Note that the proportion of overlapping users across domains are usually a very small number~\cite{cikm19:sscdr}. This data sparsity problem is further aggravated with  review scarcity, that the user documents contain incomplete and short reviews~\cite{cikm18:parl}.

To overcome these limitations, we choose to make full use of the interactions of similar non-overlapping users. We extract auxiliary reviews from \textit{like-minded} users as done in~\cite{cikm18:parl}. Specifically, for a given user-item pair, an auxiliary review is a review written by another user with the same rating score as the target user did for this item. For user $u$, her auxiliary document $D_{u_{aux}}$ is formed by merging the auxiliary reviews of the historical items purchased by user $u$ in the same domain. Note that, we only consider the auxiliary reviews from non-overlapping users, which could increase the diversity of the training data. With this data augmentation strategy, our model can still be optimized in a good shape, even when the overlapping users are very few.
	
One natural way to exploit the auxiliary document is to follow the same aspect extraction process and simply merge it with $\mathbf{A}_{u}$. However, this kind of solution ignores the fact that an auxiliary document is formed by different users who would have different language styles and different preference focuses with target user, thus may result in incompatible features. Reported in~\cite{tois19:carl}, stacking a CNN network on top of the contextual matrix is effective on rating prediction, especially when the semantics in the document are incoherent. Hence, on top of \textit{Text Convolution} used in the previous aspect extraction process, we add another convolutional layer in processing auxiliary documents, as shown in Figure~\ref{fig:architecture}.
	\begin{equation}
		c_{h,u_{aux}}^i = ReLU(\mathbf{W}_{aux}^i * \mathbf{H}_{u_{aux}}[h-\frac{s-1}{2}:h+\frac{s-1}{2}] + b^i_{aux})
	\end{equation}
	where $*$ is the convolution operator, $\mathbf{W}_{aux}^i \in \mathbb{R}^{s \times n} $ is the convolution weight matrix, $b^i_{aux}$ is the bias term, and $\mathbf{H}_{u_{aux}}$ is the feature matrix extracted by \textit{Text Convolution} in Section~\ref{ssec:aspectExt}. Similarly, we form the abstract feature matrix $\mathbf{C}_{u_{aux}} = [\mathbf{c}_{1,u_{aux}}, \mathbf{c}_{2,u_{aux}}, \ldots ,\mathbf{c}_{l,u_{aux}}]$, where $\mathbf{c}_{j,u_{aux}} \in \mathbb{R}^n$. The same \textit{Aspect Gate Control} and \textit{Aspect Attention} processes are conducted to get aspect matrices $\mathbf{A}_{u_{aux}}$ from $D_{u_{aux}}$. To update $\mathbf{A}_{u}$ with $\mathbf{A}_{u_{aux}}$ effectively, we adopt a gate mechanism based on the element-wise interactions of the corresponding aspects:
	\begin{align}
	\mathbf{g}_{aux} &= \sigma(\mathbf{W}_{f}^1[(\mathbf{A}_{u} - \mathbf{A}_{u_{aux}}) \oplus (\mathbf{A}_{u} \odot \mathbf{A}_{u_{aux}})]+\mathbf{b}_{f}^1), \\
	\mathbf{A}_{u} &= tanh(\mathbf{W}_{f}^2[\mathbf{A}_{u} \oplus (\mathbf{g}_{aux} \odot \mathbf{A}_{u_{aux}})] + \mathbf{b}_{f}^2)
	\end{align}
	where $\oplus$ is the concatenation operation, $\mathbf{W}_{f}^1, \mathbf{W}_{f}^2 \in \mathbb{R}^{k \times 2k}$ are transform matrices,  $\mathbf{b}_{f}^1, \mathbf{b}_{f}^2 \in \mathbb{R}^k$ are the bias vectors. The aspect representation $\mathbf{A}_{u}$ is updated to better profile user $u$.	
	
	\subsection{Cross-Domain Aspect Correlation Learning}
	Now, we have abstract aspect features $\mathbf{A}_{u}$ and $\mathbf{A}_{i}$ for user $u$ and item $i$ respectively. Intuitively, the rating prediction could be the aggregation of the semantic matchings between two aspects in $\mathbf{A}_{u}$ and $\mathbf{A}_{i}$ respectively. However, the matching scores would only reflect the semantic relatedness between two aspects for the specific user-item pair. Because not all aspect pairs are equally important, it is beneficial to identify global cross-domain aspect correlations. Then we can highlight the important aspect pairs at domain level for better rating prediction. To this end, we design a simple but effective method for cross-domain preference matching. Recall that we utilize a set of global aspect representations $\mathbf{V}_s$ and $\mathbf{V}_t$ to guide the aspect extraction. Here, we utilize these static aspect representations to calculate the global cross-domain aspect correlation matrix $\mathbf{S}$ as follow:
	\begin{align}
		\mathbf{S} = LeakyReLU(\mathbf{V}_{s}^{\top}\mathbf{W}\mathbf{V}_{t})\label{eqn:globalCorr}
	\end{align}
	where $\mathbf{S}(p,q)$ reflects the importance of preference transfer based on aspect $p$ from the source domain and aspect $q$ from the target domain. $\mathbf{S} \in \mathbb{R}^{M \times M}$, $\mathbf{W} \in \mathbb{R}^{k \times k}$ is a learnable matrix for affinity projection. The \textit{LeakyReLU} activation function is adopted to support the sparse aspect correlation across domains by setting the corresponding $\alpha$ to be a very small value (\eg $0.01$).

	We then calculate the semantic matching between each aspect pair between $\mathbf{A}_{u}$ and $\mathbf{A}_{i}$ as follows:
	\begin{align}
		\mathbf{S}_{u,i} &= \mathbf{A}_u^{\top}\mathbf{W}\mathbf{A}_{i}
	\end{align}
	Similar to Equation~\ref{eqn:globalCorr}, $\mathbf{S}_{u,i}(p,q)$ reflects the matching degree between the corresponding aspects; $\mathbf{W}$ is shared for affinity projection. At last, we utilize $\mathbf{S}$ as the attention weights to aggregate the pair-wise aspect matchings as the final rating prediction. 	
	\begin{align}
		\mathbf{S}^{r}_{u,i} &= \mathbf{S} \odot \mathbf{S}_{u,i} \\
		\hat{r}_{u,i} &= \frac{1}{M*M}\sum_{p=1}^{M}\sum_{q=1}^{M}{\mathbf{S}^{r}_{u,i}(p,q)} + b_u + b_{i}
	\end{align}
	Here, $b_u$ and $b_{i}$ are the user bias and item bias respectively.

\eat{
	should be determined by the static aspect correlation and the dynamic aspect connections. Specifically, the former refer to the global correlations across domains(\eg \textit{books} domain and \textit{movies} domain both have \textit{plot} aspect for transfer); the latter refer to the specific aspect-pair correlations between $\mathbf{A}_{u_s}^{merge}$ and $\mathbf{A}_{i_t}$.
}
	
	\subsection{Optimization Strategy}
	For model training, we utilize the interactions made by the overlapping users in source and target domains  for parameter optimization. Let $\mathcal{O}_s$ or $\mathcal{O}_t$ be a batch of observed user-item rating pairs in $\mathcal{D}_s$ or $\mathcal{D}_t$ respectively, restricted to $U^o$ only. The loss function of $\mathcal{L}_s$ and $\mathcal{L}_t$ can be defined as follows:
	\begin{align}
		\mathcal{L}_{\ast} &= \frac{1}{|\mathcal{O}_{\ast}|}\sum_{(u,i)\in \mathcal{O}_{\ast}}(r_{u,i}-\hat{r}_{u,i})^2 + \lambda ||\Theta_{\ast}||
	\end{align}
	where symbol $\ast$ could refer to $s$ or $t$, $\lambda$ is the regularization coefficient, and $\Theta_*$ are the trainable parameters. The two learning flows (\ie predicting $r_{u,i}$ in target domain by using $D_u$ in source domain and $D_i$ in target domain, and predicting $r_{u,i}$ in source domain by using $D_u$ in target domain and $D_i$ in source domain) are performed in turn batch after batch. Each training batch is composed of shuffled $\mathcal{O}_s$ and $\mathcal{O}_t$ at a fixed proportion, w.r.t. $|\mathcal{O}_s|/|\mathcal{O}_t| = |R_s|/|R_t|$, where $|R_s|$ and $|R_t|$ denote the number of ratings made by $U^o$ in $\mathcal{D}_s$ and $\mathcal{D}_t$ respectively. We adopt $Adam$ as the optimizer to update the parameters.
	
\section{Experiments}

	\begin{table}
		\centering
		\caption{Statistics of the three datasets in \textit{Amazon}.}
		\setlength{\tabcolsep}{1mm}{
			\scalebox{0.9}{
				\begin{tabular}{l|rrrr}
					\toprule
					Dataset & \#users & \#items & \#ratings & density \\
					\midrule
					\textit{Book} & 126,666 & 63,202 & 3,494,976 & 0.044\% \\
					\textit{Movie} (Movies and TV) & 27,822 & 12,287 & 779,376 & 0.228\% \\
					\textit{Music} (CDs and Vinyl) & 11,053 & 7,710 & 296,188 & 0.348\% \\
					\bottomrule 
		\end{tabular}}}\label{tab:stats}
	\end{table}

	\subsection{Datasets}
To evaluate our model against state-of-the-art baselines, we conduct experiments on the \textit{Amazon} review dataset~\cite{www16:dataset_amazon}. Among the largest categories,\footnote{\url{http://jmcauley.ucsd.edu/data/amazon/}}  we choose three relevant ones as three domains, namely, \textit{Book}, \textit{Movie} (named ``Movies and TV'' in Amazon) and \textit{Music} (named ``CDs and Vinyl'' in Amazon). In each domain, we remove the interaction records that are without review text, then filter out the users with fewer than 10 interactions and the items with fewer than 30 interactions following earlier studies~\cite{cikm19:sscdr,www17:ncf}. The detailed statistics of each domain is reported in Table~\ref{tab:stats}.
	
	\begin{table*}
		\centering
		\caption{Statistics of the three cross-domain recommendation scenarios. $\eta$ donotes the ratio of overlapping users included in the training set.}
		\setlength{\tabcolsep}{1mm}{
			\scalebox{0.9}{
				\begin{tabular}{lcccc|rrrrrrr}
					\toprule
					Scenario & domain & dataset & \#overlap. users & overlap. users ratio & $\eta$=100\% & $\eta$=50\% & $\eta$=20\% & $\eta$=10\% & $\eta$=5\% & \#vali. users & \#test users \\
					\midrule
					\multirow{2}*{Scenario 1} & $\mathcal{D}_s$ & \textit{Book} & \multirow{2}*{6,074} & $4.795\%$ & \multirow{2}*{3,037} & \multirow{2}*{1,518} & \multirow{2}*{607} & \multirow{2}*{303} & \multirow{2}*{151} & \multirow{2}*{1,214} & \multirow{2}*{1,823} \\
					~ & $\mathcal{D}_t$ & \textit{Movie} & ~ & $21.832\%$ & ~ & ~ & ~ & ~ & ~ \\
					\midrule
					\multirow{2}*{Scenario 2} & $\mathcal{D}_s$ & \textit{Movie} & \multirow{2}*{2,782} & $9.999\%$ & \multirow{2}*{1,391} & \multirow{2}*{695} & \multirow{2}*{278} & \multirow{2}*{139} & \multirow{2}*{69} & \multirow{2}*{556} & \multirow{2}*{835}\\
					~ & $\mathcal{D}_t$ & \textit{Music} & ~ & $25.170\%$ & ~ & ~ & ~ & ~ & ~ \\
					\midrule
					\multirow{2}*{Scenario 3} & $\mathcal{D}_s$ & \textit{Book} & \multirow{2}*{1,705} & $1.346\%$ & \multirow{2}*{853} & \multirow{2}*{426} & \multirow{2}*{170} & \multirow{2}*{85} & \multirow{2}*{42} & \multirow{2}*{340} & \multirow{2}*{512}\\
					~ & $\mathcal{D}_t$ & \textit{Music} & ~ & $15.426\%$ & ~ & ~ & ~ & ~ & ~ \\
					\bottomrule
\label{tab:cr-stats}
		\end{tabular}}}
	\end{table*}
	
As the three domains are relevant to each other,  we construct three cross-domain scenarios in pairs. In each scenario, we choose the domain with more users as $\mathcal{D}_s$ and the other as $\mathcal{D}_t$. Following the settings in~\cite{cikm19:sscdr}, we randomly sample 50\% of the overlapping users to be cold-start users, \ie  their interactions in $\mathcal{D}_t$ are not seen by the models, but are used for validation and testing purposes (specifically, 30\% are set for test users and 20\% are set for validation users). The remaining 50\% of overlapping users are used for training purpose. In order to simulate different ratios of overlapping users, we building our training set by randomly including a certain fraction $\eta \in \{100\%, 50\%, 20\%, 10\%, 5\%\}$ of the remaining 50\% overlapping users. The detailed statistics of each cross-domain scenario is reported in Table~\ref{tab:cr-stats}.

	\subsection{Baseline Methods}
	We compare against the following baselines, including the traditional ones and recent state-of-the-arts.
	\begin{itemize}[leftmargin=1.8em]
		\item \textbf{CMF}~\cite{kdd08:cmf} is a simple and well-known method for cross-domain recommendation by sharing the user factors and factorizing joint rating matrix across domains.
		
		\item \textbf{EMCDR}~\cite{ijcai17:emcdr} is the first to propose the three-step optimization paradigm by training matrix factorization in both domains successively then utilizing multi-layer perceptrons to map the user latent factors.
		
		\item \textbf{CDLFM}~\cite{dasfaa18:cdlfm} modifies matrix factorization by fusing three kinds of user similarities as a regularization term based on their rating behaviors.  A neighborhood-based mapping approach is used to replace the previous multi-layer perceptrons, by considering similar users and the gradient boosting trees (GBT) based ensemble learning method.
		
		\item \textbf{DFM}~\cite{aaai19:rc_dfm} is a simple version of RC-DFM~\cite{aaai19:rc_dfm}. It leverages the work of aSDAE~\cite{aaai17:asdae} to generate user representations from rating matrix, with multi-layer perceptrons to conduct mapping as well.
		
		\item \textbf{R-DFM}~\cite{aaai19:rc_dfm} is another variant of RC-DFM~\cite{aaai19:rc_dfm}\footnote{In the RC-DFM paper, the item content is fused into another aSDAE to let it close to its review-based representation. However, the authors do not explain the detailed method to obtain item content. Moreover, the improvement of RC-DFM over R-DFM is very small, so we choose R-DFM for comparison.}. It combines the rating records and the reviews by an extended aSDAE to enhance the user/item representations. The mapping part is also multi-layer perceptrons.
		
		\item \textbf{ANR}~\cite{cikm18:anr} is a state-of-the-art review-based single-domain approach by performing aspect matching for user-item pair. Here we conduct recommendation by leveraging their corresponding reviews in the source domain directly, after training the model purely on the target domain.
	\end{itemize}

	\subsection{Experimental Setup}	
	We preprocess both user and item documents in all datasets, following the related studies~\cite{recsys16:cmf4dcr,tois19:carl}: 1) remove stop words and words with high document frequency  (\ie relative document frequency above 0.5); 2) choose the top 20,000 words as vocabulary according to their tf-idf score and remove other words from the raw documents; 3) amputate (pad) the long (short) documents to the same length of $500$ words. We utilize the 300-dimension word embeddings pre-trained in Google News\footnote{\url{https://code.google.com/archive/p/word2vec/}}~\cite{nips13:word2vec} to get the initial embedding vector for each word.
	
	We apply grid search to tune the hyper-parameters for all the methods based on the setting strategies reported in their papers. The final performances of all methods are reported over 5 runs.
	
	For CATN\footnote{Our implementation is available at \url{https://github.com/AkiraZC/CATN}}, the number of convolution filters $n$ = 50, window size $s$ = 3. The batch size (number of $\mathcal{O}_s \cup \mathcal{O}_t$) is 256. The dropout strategy is applied to ignore a small percent of values in aspect representations randomly during the training process. The keep probability of dropout is set to be $0.8$, and we choose learning rate to 0.001 for model training. The latent dimension size $k$ is optimized from \{16, 32, 64, 128\}, and the aspect number $M$ is optimized from \{3, 5, 7, 9\}.
	
	For evaluation metric, we use \textit{MSE} as performance metric, which is widely adopted in many related works for performance evaluation~\cite{cikm18:anr,kdd18:mpcn,sigir19:carp,tois19:carl}, formulated as: 	
	$$\mathit{MSE}=
	\frac{1}{|\mathcal{O}|}\sum_{(u,i)\in \mathcal{O}}(r_{u,i}-\hat{r}_{u,i})^2$$
	where $\mathcal{O}$ is the cold-start user validation set for parameter selection or test set for performance comparison.
	
	\begin{table*}
		\centering
		\caption{Performance comparison on three recommendation scenarios in terms of \textit{MSE}. The best and second best results are highlighted in boldface and underlined respectively. $\blacktriangle\%$ denotes the relative improvement of CATN over the best SOTA algorithm. All reported improvements over baseline methods are statistically significant at a 0.05 level.}
		\label{tab:results}
		\setlength{\tabcolsep}{1mm}{
			\scalebox{0.9}{
				\begin{tabular}{c||ccccc|ccccc|ccccc}
					\toprule[1.2pt]
					{Scenario} & \multicolumn{5}{c|}{Scenario 1} & \multicolumn{5}{c|}{Scenario 2} & \multicolumn{5}{c}{Scenario 3} \\ \midrule
					{$\mathcal{D}_s \rightarrow \mathcal{D}_t$} & \multicolumn{5}{c|}{\textit{Book} $\rightarrow$ \textit{Movie}} & \multicolumn{5}{c|}{\textit{Movie}  $\rightarrow$ \textit{Music}} & \multicolumn{5}{c}{\textit{Book}  $\rightarrow$ \textit{Music}} \\ \midrule
					\diagbox{Method}{$\eta$} & 100\% & 50\% & 20\% & 10\% & 5\% & 100\% & 50\% & 20\% & 10\% & 5\% & 100\% & 50\% & 20\% & 10\% & 5\% \\ \midrule[1pt]
					CMF & 1.167 & 1.169 & 1.179 &1.179& 1.181 & 1.139 & 1.140 & 1.158 &1.167& 1.173 & 0.939 & 0.942 & 0.962 &0.967& 0.970 \\
					EMCDR & 1.129 & 1.138 & 1.142 &1.140& 1.148 & 1.116 & 1.138 & 1.144 &1.172& 1.175 & 0.924 & 0.927 & 0.934 &0.936& \underline{0.937} \\
					CDLFM & 1.126 & 1.130 & 1.135 & 1.138 & 1.144 & 1.115 & 1.133 & 1.145 & 1.169 & 1.171 & 0.918 & 0.925 & 0.930 & 0.931 & 0.951 \\
					DFM & 1.141 & 1.143 & 1.149 & 1.150 & 1.156 & 1.136 & 1.158 & 1.162 & 1.166 & 1.175 & 0.923 & 0.929 & 0.933 & 0.941 & 0.952 \\
					R-DFM & 1.132 & 1.135 & 1.141 & 1.146 & 1.152 & 1.128 & 1.143 & 1.146 & \underline{1.150} & 1.166 & 0.911 & 0.917 & 0.928 & 0.936 & 0.943  \\
					ANR & \underline{1.123} & \underline{1.127} & \underline{1.130} & \underline{1.135} & \underline{1.137} & \underline{1.122} & \underline{1.137} & \underline{1.142} & 1.155 & \underline{1.160} & \underline{0.895} & \underline{0.903} & \underline{0.912} & \underline{0.919} & 0.940 \\
					CATN & \textbf{1.049} & \textbf{1.072} & \textbf{1.079} &\textbf{1.093} & \textbf{1.097} & \textbf{1.042} & \textbf{1.075} & \textbf{1.102} &\textbf{1.126} & \textbf{1.144} & \textbf{0.862} & \textbf{0.868} & \textbf{0.875} &\textbf{0.896}& \textbf{0.899} \\
					\hline
					$\blacktriangle\%$ & 6.59 & 4.88 & 4.51 & 3.70 & 3.52 & 6.55 & 5.45 & 3.50 & 2.09 & 1.38 & 3.69 & 3.88 & 4.06 & 2.50 & 4.06 \\
					\bottomrule[1.2pt]
		\end{tabular}}}
	\end{table*}

\begin{figure*}
	\centering
		\includegraphics[width=0.90\linewidth]{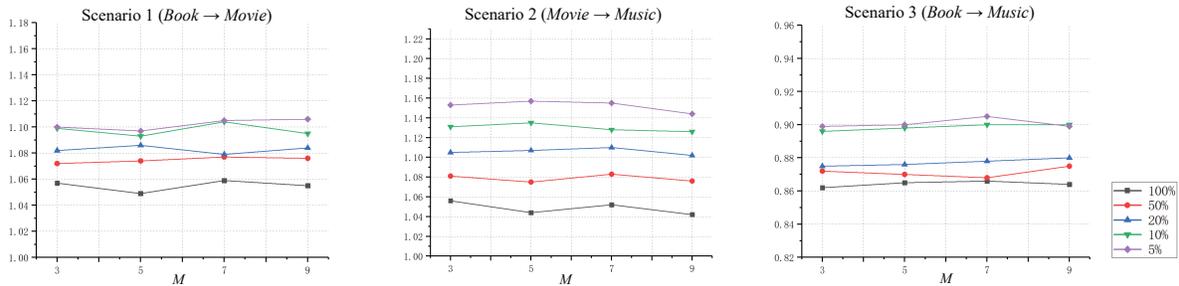}
		\caption{Impact of number of aspects  $M$ in CATN. }
\label{fig:M}
	\end{figure*}
	
	\subsection{Results and Discussion}
	The overall results of all methods over the three cross-domain recommendation scenarios  are
	reported in Table~\ref{tab:results}. We made the following observations from the results.

First of all, CATN outperforms all baselines significantly on all cross-domain recommendations, and in terms of different ratios of overlapping users in all settings. This result  demonstrates the superiority of our review-based recommendation for cold-start users in cross-domain setting.
		
It's no surprise that CMF consistently yields the worst performance on all evaluations. Learning user representations simply by factorizing a joint matrix is not adequate, which is also consistent with what has been observed in earlier studies~\cite{ijcai17:emcdr,dasfaa18:cdlfm,aaai19:rc_dfm}.  CDLFM makes some improvements to the user factors learning and the cross-domain mapping processes, which leads to conspicuous improvements over EMCDR. R-DFM modifies DFM ulteriorly by fusing user reviews. However, none of them ever achieves the best result, which verifies the drawbacks of the straightforward three optimization process as shown in Figure~\ref{fig:3-step}.

For DFM and R-DFM, according to our experiments, the results suffer from declination compared to EMCDR. This is because, aSDAE takes original rating vectors as input, which can be over a hundred of thousand dimensions in our dataset. In this case, millions of training parameters need to be optimized, which makes the model rather complicated to converge\footnote{To avoid the issue of parameters explosion, the authors of~\cite{aaai19:rc_dfm} preprocess the data with the \textit{120-cores} settings (\ie filter out the items with fewer than $120$ interactions). However, it does not fit real-world sparse recommender scenarios.} and yield inferior results. Although ANR is not designed for cross-domain scenarios, it maintains competitive results over the other baselines, confirming the usefulness of review information for the recommendation task.

From the results, we observe that while the methods based on three-step optimization are sensitive to $\eta$, especially when the ratio is low (10\% or 5\%), our CATN shows more robust performance. As $\eta$ gets lower, the overlapping users get fewer. Existing cross-domain mapping cannot be well trained  because of  the lack of training instances, resulting in an inferior result. On the contrary, CATN utilizes a simple but effective way to emphasizes the transfer of cross-domain aspects, instead of the user representations directly. In this way, CATN reduces the impact by $\eta$ to a large extent.

	\section{Model Analysis}
We now present detailed analysis of the proposed CATN model.  We first investigate the impact of  hyper-parameter settings (\ie $M$) to the performance of CATN. Next, we conduct three ablation studies to analyze how different components in our proposed model contribute to the overall results. Lastly, study cases are shown to give explainable analysis of the cross-domain aspect transfer process.

	\begin{table*}
		\centering
		\caption{Performance comparison of the three model variants on three recommendation scenarios.}
		\label{tab:ablation}
		\setlength{\tabcolsep}{1mm}{
			\scalebox{0.9}{
				\begin{tabular}{c||ccccc|ccccc|ccccc}
					\toprule[1.2pt]
					{Scenario} & \multicolumn{5}{c|}{Scenario 1} & \multicolumn{5}{c|}{Scenario 2} & \multicolumn{5}{c}{Scenario 3} \\ \midrule
					{$\mathcal{D}_s \rightarrow \mathcal{D}_t$} & \multicolumn{5}{c|}{\textit{Book} $\rightarrow$ \textit{Movie}} & \multicolumn{5}{c|}{\textit{Movie}  $\rightarrow$ \textit{Music}} & \multicolumn{5}{c}{\textit{Book}  $\rightarrow$ \textit{Music}} \\ \midrule
					\diagbox{Method}{$\eta$} & 100\% & 50\% & 20\% & 10\% & 5\% & 100\% & 50\% & 20\% & 10\% & 5\% & 100\% & 50\% & 20\% & 10\% & 5\% \\ \midrule[1pt]
					CATN-basic & 1.103 & 1.109 & 1.117 & 1.122 & 1.127 & 1.114 & 1.127 & 1.144 & 1.158 & 1.160 & 0.881 & 0.889 & 0.897 & 0.900 & 0.903 \\
					CATN-attn & 1.084 & 1.102 & 1.109 & 1.116 & 1.121 & 1.074 & 1.103 & 1.131 & 1.153 & 1.157 & 0.880 & 0.893 & 0.895 & 0.899 & 0.901 \\
					CATN-separate & 1.056 & 1.079 & 1.087 & 1.099 & 1.103 & 1.055 & 1.085 & 1.115 & 1.137 & 1.153 & 0.868 & 0.870 & 0.884 & 0.899 & \textbf{0.899} \\
					CATN & \textbf{1.049} & \textbf{1.072} & \textbf{1.079} & \textbf{1.093} & \textbf{1.097} & \textbf{1.042} & \textbf{1.075} & \textbf{1.102} & \textbf{1.126} & \textbf{1.144} & \textbf{0.862} & \textbf{0.868} & \textbf{0.875} & \textbf{0.896} & \textbf{0.899} \\
					\bottomrule[1.2pt]
		\end{tabular}}}
		\vspace{-0.cm}
	\end{table*}
	
	\subsection{Aspect Number Sensitivity}
	Figure~\ref{fig:M} plots the effect of varying $M \in  \{3, 5, 7, 9\}$ for CATN across multiple evaluations settings, with different preset $\eta$ values.
In general, a small $M$ leads to coarse aspects, while a large $M$ leads to fine-grained aspects. However, as we discussed earlier, not all aspects from source and target domains would match and participate in the preference transfer, and the attention mechanism would learn optimal weights between matching aspects. In this sense, the varying of $M$ would only affect the number of aspects in source and target domains and does not affect much of the preference transfer. As shown in the plot, given the same setting (\ie a fixed  $\eta$ in a particular cross-domain recommendation task), the performance fluctuations incurred by different $M$ values are very small, suggesting that CATN is robust to this parameter setting.

On the other hand,  the setting of  $\eta$ directly affects the number of overlapping users from whom the system learns the preference matching across domains. It is clear that more overlapping users lead to a better understanding of preferences across domains, hence better recommendation accuracy.
	
	\subsection{Ablation Study}
	Reflecting the intuition of CATN, we design a cross-domain review-based preference matching procedure with two learning flows. The learning process involves  global-sharing aspect representations $\mathbf{V}_s$ and $\mathbf{V}_t$ to guide the aspect extraction. The global cross-domain aspect correlations $\mathbf{S}$ are exploited to give final predictions. In addition, auxiliary reviews from like-minded and non-overlapping users are exploited to enhance user aspect extraction, with the aim of alleviating the data sparsity issue. Accordingly, we come up with three variants of CATN as follows:
	\begin{itemize}[leftmargin=1.8em]
		\item \textbf{CATN-basic}: As the basic variant of CATN, it shares aspect extraction parameters in the two learning flows. We exclude $\mathbf{V}_s$ and $\mathbf{V}_t$ by replacing the attention mechanism with a \textbf{simple average} operation in Equation~\ref{eqn:aspect}. The prediction only considers the aspect matchings. No auxiliary reviews are exploited in this variant.
		\item \textbf{CATN-attn}: In contrast to CATN-basic, we introduce the global-sharing aspect representations to fulfill aspect extraction, and the global cross-domain aspect correlations are taken into account. In other words, CATN-attn is a simplified version of CATN without domain-specific aspect extraction and auxiliary reviews.
		\item \textbf{CATN-separate}: In contrast to CATN-attn, we leverage two separate aspect extraction parameters in the two learning flows. In other words, CATN-separate is a simplified version of CATN by not including auxiliary reviews.
	\end{itemize}

	The results of the ablation studies on all evaluation settings are reported in Table~\ref{tab:ablation}. We make the following observations: 1) With reference to the results in Table~\ref{tab:results},  CATN-basic outperforms most baselines in all recommendation scenarios, demonstrating the effectiveness of the cross-domain aspect-based transfer approach; 2) CATN-attn gains some improvements over CATN-basic, which reveals the benefit of including global-sharing aspect representations; 3) CATN-separate outperforms the above variants, which shows the usefulness of distinct aspect extractions; and 4) As the integrated model, CATN improves the performance further by exploiting auxiliary reviews from like-minded and non-overlapping users. This observation suggests that auxiliary reviews are of vital value to alleviate the data sparsity issue.
	
	\subsection{Optimization Efficiency}
	Designed as an end-to-end learning framework, our proposed CATN not only overcomes the deficient three-step optimization process, but also speeds up the training time by optimizing only the ratings from overlapping users.
	
	Specifically, in existing approaches, the third-step cross-domain transfer process cannot be conducted until the first two steps reach their optimal states, which is time-consuming. Besides, DFM and R-DFM contain massive parameters in terms of their Auto-Encoder component, thus hindering the convergence speed.
	
	In our experiments, CATN spends about $600$s (second) to reach the best validation performance in \textit{Book} $\rightarrow$ \textit{Movie} at $\eta=50\%$, by using one Nvidia 1080 GPU. In contrast, it is $300$s for CMF, $400$s for ANR, $1000$s for EMCDR, $1200$s for CDLFM, over 1 hour for DFM and R-DFM. While our CATN achieves the best performance, it maintains a competitive training time over the other baselines, especially in terms of the review-based approach R-DFM.
		
	\begin{figure}
		\centering
		\includegraphics[width=0.98\linewidth]{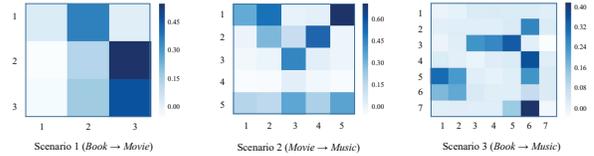}
		\caption{Global aspect correlation matrix $S$ on three recommendation scenarios at $\eta=50\%$.}\label{fig:S_attn}
	\end{figure}

    \begin{table*}
		\scriptsize
		\centering
		\caption{Example study of three user-item pairs from three recommendation scenarios at $\eta=50\%$.}
		\begin{tabular}{c|p{12cm}}
			\toprule
			\multicolumn{2}{c}{scenario 1 (\textit{Book} $\rightarrow$ \textit{Movie}): $r_{u,i}=5.0 ,\hat{r}_{u,i}=4.72 $} \\
			\midrule
			$A_{u}[2]$ &...I enjoyed reading the book. It adds a lot to the movie. I think the \hlgrn{biggest plot element} is that it really expands upon the Garthim-Master's \hlgrn{character}...\\
			\midrule
			$A_{u_{aux}}[2]$ & ...This is \hlgrn{an interesting history}. Much of the book is \hlgrn{interesting and readable}...\\
			\midrule
			$A_{i}[3]$ & ...all the \hlgrn{characters} playerd their \hlgrn{roles} well and overall, it was a \hlgrn{fun} movie to watch...\\
			\midrule
			$A_{u}[3]$ in $D_u^t$ & ..I felt this movie was more than just a fluffy \hlgrn{romantic comedy}. I was pleasantly surprised when...\\
			\midrule
			target review $d_{u,i}$ & ...like a double \ulred{romantic comedy}...all of the \ulred{actors are wonderful}. Andy Griffith is absolutely charming as the "player" grandfather...\\
			\midrule \midrule
			
			\multicolumn{2}{c}{scenario 2 (\textit{Movie} $\rightarrow$ \textit{Music}): $r_{u, i}=3.0 ,\hat{r}_{u, i} =3.58 $} \\
			\midrule
			$A_{u}[1]$ & ...I think all criticism aside, this is a very \hlgrn{interesting and engrossing} film...\\
			\midrule
			$A_{u_{aux}}[1]$ & ...\hlgrn{tender plots} are always my favorite...\\
			\midrule
			$A_{i}[5]$ & ...I like sting. And I like some \hlgrn{renaissance} music...\hlgrn{lute} plus sting sounded like a good idea, but about half of one...\\
			\midrule
			$A_{u}[5]$ in $D_u^t$ & ...I appreciate the \hlgrn{technical merits} of the membership...\\
			\midrule
			target review $d_{u, i}$& ...As much as I appreciate when an artist stretches his ability, Sting should have reconsidered his \ulred{venture into lute music}...but enventually repetitive and unoriginal...\\
			\midrule \midrule
			
			\multicolumn{2}{c}{scenario 3 (\textit{Book} $\rightarrow$ \textit{Music}): $r_{u, i}=4.0 ,\hat{r}_{u, i} =4.33 $} \\
			\midrule
			$A_{u}[7]$ & ...This is a fairly entertaining novel that \hlgrn{reads more like} a collection of loosely connected \hlgrn{stories} held together by the common thread of...\\
			\midrule
			$A_{u_{aux}}[7]$ & ...is the \hlgrn{most grossly misogynist and sexist novel} I've read in a long time which is distressing considering...\\
			\midrule
			$A_{i}[6]$ & ...because it's such a \hlgrn{lovely, upbeat optimistic pop song}...what a pity that lyrics are just slightly above alternative teenage-pop...\\
			\midrule
			$A_{u}[6]$ in $D_u^t$ & ...I've been a fan of \hlgrn{trip hop} since I heard blue lines from massive attack and I've heard many bad imitations since then... \\
			\midrule
			target review $d_{u, i}$ & ...overall it pleasantly comes together as a whole creating a lovely collection of what I could call \ulred{downtempo triphop pop}... \\
			\bottomrule
		\end{tabular} \label{tab:showcase}
	\end{table*}	

	\begin{table*}
		\normalsize
		\centering
		\caption{Top-5 words for each aspect in \textit{Example 1}. The ``Aspect Labels'' are manually generated based on our own interpretation.}
		\label{tab:aspect_words}
		\setlength{\tabcolsep}{1mm}{
			\scalebox{0.9}{
				\begin{tabular}{c|ccc|ccc|ccc}
					\toprule
					Aspect & $A_{u}[1]$ & $A_{u}[2]$ & $A_{u}[3]$ & $A_{u_{aux}}[1]$ & $A_{u_{aux}}[2]$ & $A_{u_{aux}}[3]$ & $A_{i}[1]$ & $A_{i}[2]$ & $A_{i}[3]$ \\
					\midrule
					Aspect Label & \textbf{Writing} & \textbf{Plot} & \textbf{Scene} & \textbf{Writing} & \textbf{Plot} & \textbf{Scene} & \textbf{Acting} & \textbf{Feeling} & \textbf{Content} \\
					\midrule
					\multirow{5}*{Top-5 Words} & dialogue & plot & recommend & mislead & interesting & exciting & truly & hilarious & role \\
					& texture & charactor & story & plausible & culture & generic & playing & entertaining & classic \\
					& complex & write & landscape & plot & readable & actions & fan & fun & character \\
					& weave & stereotype & eye & possibilities & species & imagination & wisdom & charismatic & funny \\
					& nuanced & expedition & looking & occurred & plot & title & chasing & enjoy & comedy \\
					\bottomrule
		\end{tabular}}}
	\end{table*}
		
\subsection{Explainability Analysis}
We further investigate whether CATN can discover meaningful aspect transfers across domains. To better visualize the aspects, we retrieve the top-5 words whose weights are the average of attention scores (\ie $\beta_{m,j,u}$ and $\beta_{m,j,i}$) in the document. Then we display the sentences containing these informative words for better understanding.

The user-item pair from each cross-domain recommendation setting at $\eta=50\%$ is randomly sampled and displayed in Table~\ref{tab:showcase} to offer semantic explanations.  As Figure~\ref{fig:S_attn} illustrates, the global aspect correlations across domains are extremely sparse (usually focusing on one particular block).  Due to space limitation, we pick the \textbf{most correlated} aspects-pair across domains (\ie corresponding to the maximum value in the matrix $\mathbf{S}$), and list the aspects extracted from user document, user auxiliary document of $\mathcal{D}_s$ and item document of $\mathcal{D}_t$ respectively. We also list the aspect information mentioned in the corresponding user document (namely $D_u^t$) in $\mathcal{D}_t$. The informative phrases are highlighted in orange color, including the stop words inside the context. As a reference, we underline the corresponding parts in the target review $d_{u,i}$ using red lines.

\paratitle{Example 1}: The first example shows the aspect transfer process from \textit{Book} domain to \textit{Movie} domain. For better explanation, we list the top-5 words of each aspect in Table~\ref{tab:aspect_words} and summarize the ``Aspect Labels'' based on our examination. We observe that the second and the third aspect of \textit{Book} (\ie \textit{plot}, \textit{scene}) are the most closely related to the third aspect of \textit{Movie} (\ie \textit{content}). This  is consistent with what is shown in Figure~\ref{fig:S_attn}.
	
In detail, from $A_{u}[2]$ and $A_{u_{aux}}[2]$ we can infer that the user likes interesting plots or stories. According to $A_{i}[3]$, this item is described as a funny movie with well-played roles. Hence the preferred aspect of the user can be transferred reasonably, leading to a high prediction accuracy. Note that, the information mentioned in $D_u^t$ for $A_{u}[3]$ and the indication of user likes comedy in the target review both suggest the correctness of aspect transfer.
	
\paratitle{Example 2}: The second example shows the aspect transfer from \textit{Movie} domain to \textit{Music} domain. According to Figure~\ref{fig:S_attn}, the most related aspect-pair are the first aspect of $\mathcal{D}_s$ and the fifth aspect of $\mathcal{D}_t$. Correspondingly, from $A_{u}[1]$ and $A_{u_{aux}}[1]$ we can infer that this user prefers peaceful and tender plots in movies. According to $A_{i}[5]$, this item is indeed described as a soothing lute music with some negative comments. In view of the preference of the user and the reputation of the item, our model gives a moderate-score prediction, which turns out to be accurate in reference to the target review.
	
\paratitle{Example 3}: The third example shows the aspect transfer  from \textit{Book} domain to \textit{Music} domain. Similarly, according to Figure~\ref{fig:S_attn}, the most related aspect pair is the seventh aspect of \textit{Book} and the sixth aspect of \textit{Music}. From $A_{u}[7]$ and $A_{u_{aux}}[7]$ we can infer that this user is fond of beautiful words and stories. According to $A_{i}[6]$, this item is a pop song praised for its melody but criticized for its lyrics. Considering the user's preference for stories, our CATN still gives a high-score prediction in term of the item's characteristics and aspect transfer tendency.
	
Overall, the three sets of examples show that CATN is effective  in cross-domain recommendation for cold-start user, with reasonable aspect transfer to support semantic explanation. 
	\section{Conclusion}

In this paper, we study the problem of review-based cross-domain recommendation for cold-start users. Our key focus is on the transfer of user preference derived from source domain to target domain, for effective and explainable recommendation. Instead of following the existing framework to first learn user/item representations in the source and target domains, then learn the mapping, we propose an end-to-end learning strategy. More importantly, we consider the fact that users' preferences are multi-faceted and only a subset of aspects in the two domains would match. In our framework, we therefore derive aspects from review documents and aim to find their correlations through a global aspect representations with attention. We show that our CATN model outperforms all existing models for cross-domain recommendation tasks. We believe that CATN offers an alternative view of the this interesting and critical task. Our study would also trigger studies on more effective ways of modeling user preference transfer across different domains. Inspired by~\cite{sigir20:lightgcn}, we may investigate for more possibilities on graph-based CDR in the future.
	
	\begin{acks}
		This work was supported by Alibaba Group through Alibaba Innovative Research Program and National Natural Science Foundation of China (No.~61872278). Chenliang Li is the corresponding author.
	\end{acks}
	\bibliographystyle{ACM-Reference-Format}
	\bibliography{acmart}
\end{document}